\documentclass[aps,prd,nofootinbib,superscriptaddress,twocolumn]{revtex4-1}

\usepackage{amsmath}
\usepackage{amssymb}
\usepackage{bbold}
\usepackage{epsfig}
\usepackage{graphicx}
\usepackage{color}
\usepackage{nicefrac}
\usepackage{epstopdf}
\usepackage{tabularx}
\usepackage[colorlinks=true, pdfstartview=FitV, linkcolor=blue, citecolor= red,urlcolor=blue]{hyperref}
\definecolor{darkgreen}{rgb}{0,0.5,0}
\definecolor{purple}{rgb}{0.5,0,0.5}
\definecolor{nblue}{rgb}{0.0,0.0,0.50}
\definecolor{scarlet}{rgb}{1.0,0.2,0}

\newcommand{\Tr}{\operatorname{Tr}_\mathrm{CD}}
\graphicspath{{PlotLC/}}

\begin{document}

\title{The quark gap equation in light-cone gauge}

\author{Roberto Correa da Silveira}
\affiliation{Instituto de F\'{\i}sica Te\'orica, Universidade Estadual Paulista, Rua Dr.~Bento Teobaldo Ferraz 271, 01140-070 S\~ao Paulo, S\~ao Paulo, Brazil}

\author{Fernando E. Serna}
\affiliation{Departamento de F\'isica, Universidad de Sucre, Carrera 28 No.~5-267, Barrio Puerta Roja, Sincelejo 700001, Colombia}

\author{Bruno El-Bennich}
\affiliation{Instituto de F\'{\i}sica Te\'orica, Universidade Estadual Paulista, Rua Dr.~Bento Teobaldo Ferraz 271, 01140-070 S\~ao Paulo, S\~ao Paulo, Brazil}
\affiliation{Departamento de F\'isica, Universidade Federal de S\~ao Paulo, Rua S\~ao Nicolau 210, Diadema, 09913-030 S\~ao Paulo, Brazil}


\begin{abstract}
An exploratory nonperturbative calculation of the quark propagator in light-cone gauge is motivated by  distribution amplitudes whose definition implies a Wilson line. 
The latter serves to  preserve  the gauge invariance of the hadronic amplitudes and becomes trivial in light-cone gauge. To that end, we explore the corresponding  
Dyson-Schwinger equation in its leading truncation and with a dressed vertex derived from a Ward identity in light-cone gauge. The quark's mass and wave renormalization 
functions are found  to depend on the  orientation of the quark momentum relative to the light-like four-vector below 1~GeV, which expresses the light-cone gauge 
dependence of the propagator, while a third, complex-valued amplitude exhibits little dependence on that orientation and vanishes in the quark's rest frame.
\end{abstract}

\date{\today}
\maketitle


\section{Introduction}
\label{intro}

Noncovariant gauges have a long history and, as their name implies, one of their features is the breaking of relativistic covariance. A typical example is provided by 
the light-cone gauge, which is a physical gauge defined by,
\begin{equation}
     n \cdot A^a  = 0 \, ,
  \label{lcgauge}   
\end{equation}
where $A^a_\mu (x)$ is a massless Yang-Mills field and the light-like four-vector defined by $n^2 =0$. Since the four-vector $n$ defines a preferential
axis in space-time, the above condition is more generally referred to as \emph{axial gauge\/}. Despite the lack of covariance, the strong interest in axial gauges can be 
attributed to the decoupling of the Faddeev-Popov ghosts from the gauge field, thereby eliminating the unphysical degrees of freedom in the theory~\cite{Leibbrandt:1987qv}.
This is because the ghost-gluon vertex is proportional to the vector $n$ which projects out the gluon fields due to the gauge condition in Eq.~\eqref{lcgauge}. 
Consequently, $n^\mu D_{\mu \nu}^{a b}=0$ in any Feynman diagram, where $D_{\mu \nu}^{a b}$ is the gluon propagator in light-cone gauge. This observation 
alone may be of  interest to the nonperturbative calculation of the gluon propagator~\cite{Aguilar:2006gr,Aguilar:2008xm}, as the gluon also decouples from 
the ghost in the Dyson-Schwinger equation~(DSE). 

There are other reasons to consider  light-cone gauge, in particular when facing the difficulty of evaluating nonperturbative matrix elements that contain 
a  Wilson line~\cite{Costa:2021mpk}. The phenomenological motivation stems from the definition of light-cone distribution amplitudes and parton distribution 
functions~\cite{Jaffe:1996zw}, which contain a Wilson line to preserve gauge symmetry. More precisely, there exists a body of work that addresses parton distribution 
within a DSE and Bethe-Salpeter equation  approach of the mesons in Landau gauge, assuming the effect of the Wilson line is negligible~\cite{Serna:2020txe,
Serna:2022yfp,Chang:2013pq,Shi:2015esa}. An alternative approach is to project the Bethe-Salpeter wave function on the light front and derive therefrom mesonic 
distributions functions, which avoids the Wilson line altogether but is so far limited to the leading Fock state~\cite{Mezrag:2016hnp,Shi:2018zqd,Serna:2024vpn}.

The parton distribution function of a pseudoscalar mesons is defined as,
\begin{align}
    q_M(x) = & \int \frac{ d\lambda}{4\pi}  \,  e^{-i x\lambda P\cdot n}  \nonumber  \\
              &  \times \,   \langle M | \bar \psi_q(0) \gamma \cdot n\, W (0, \lambda n)  \psi_q(\lambda n) | M \rangle  \, ,
\end{align}
where $x = k \cdot n / P \cdot n$ is the light-cone momentum fraction carried by the struck quark and the Wilson line reads,
\begin{equation} 
   W(0, n \lambda) = \mathcal{P} e^{-i g \int_\lambda^0  n\, \cdot A(n \xi)   d\xi} \, ,
\end{equation}
in which $\mathcal{P}$ denotes the path-ordering operator. In light-cone gauge~\eqref{lcgauge} this operator is trivial, which is why the calculation of distributions 
is attractive in this gauge. However, this comes at a price: the necessity to compute nonperturbative quark propagators in light-cone gauge. In a functional continuum 
approach to Quantum Chromodynamics (QCD), however, this is commonly done by solving the respective DSEs of the fermion and gauge fields in Landau or covariant 
$R_\xi$ gauges~\cite{Lessa:2022wqc}. 

In here, we take a first step to fill this gap and solve the quark DSE in light-cone gauge. We start with the general form of the quark propagator and discuss divergences 
in Feynman diagrams that stem from the denominator $(n\cdot q)^{-1}$ in the gluon propagator. We then formulate the DSE of the quark in light-cone gauge and address 
this issue within this nonperturbative framework. The solution can be decomposed into three Lorentz-invariant amplitudes, two of which are real  and play the role of the 
mass and wave function renormalization known of covariant gauges, while the third amplitude is complex. It is found that all amplitudes exhibit an angular dependence 
on the light-like vector $n$. We interpret this behavior as a \emph{geometric} expression of gauge dependence of the quark propagator in light-cone gauge.


\section{Gluon propagator in light-cone gauge}
\label{sec2}

The light-cone gauge ~\eqref{lcgauge} is implemented with a gauge fixing term restricting the degrees of freedom in the QCD Lagrangian,
\begin{equation}
   \mathcal{L}_{\text{fix}} = - \frac{1}{2 \alpha}\left(n \cdot A^a\right)^2 .
\label{gaugefix}   
\end{equation}
From the Yang-Mills Lagrangian and the gauge fixing term~\eqref{gaugefix} one can derive the Euler-Lagrange equation of a non-interacting gluon. After taking $\alpha \to 0$, 
this leads to the Green function, 
\begin{equation}
   D_{\mu \nu}^{a b}  (q) =  \frac{-i \delta^{a b}}{q^2+i \epsilon}\left(g_{\mu \nu}-\frac{n_\mu q_\nu+n_\nu q_\mu}{n \cdot q}\right) \, ,
 \label{LCgluon}  
\end{equation}
which is the gluon propagator in light-cone gauge that satisfies, 
\begin{equation}
    n^\mu D_{\mu \nu}(q)=n^\nu D_{\mu \nu}(q)=0 \, .
\end{equation}

Any Feynman diagram containing a gluon propagator~\eqref{LCgluon} also includes the  $(n\cdot q)^{-1}$ pole. Divergences arising from this pole can be dealt with using a 
principal value prescription, though an attractive alternative is offered by the Mandelstam-Leibbrandt [ML] prescription~\cite{Leibbrandt:1983pj,Leibbrandt:1987uc,Mandelstam:1982cb}.
The latter modifies the denominator by adding a small imaginary shift $i\theta$ which is taken to zero after Wick rotation. In order to do so, the denominator must be put in a form 
that allows for a Wick rotation without hitting poles. The ML prescription rests on the observation that the definition of a light-like vector with $n^2=0$ does not constrain the 
vector unambigously:
\begin{equation} 
   n_0^2 - \vec n^{\,2} = 0 \ \ \Longrightarrow \  \  n_0 = \pm |\vec n| \, . 
\end{equation}
Therefore, the location of the  $(n\cdot q)^{-1}$  pole is not unique. This ambiguity can be addressed by choosing $n = (|\vec n|, \vec n)$ and introducing a dual vector 
$n^* = (|\vec n|, -\vec n)$, $n^{*2} =0$, $n^*\cdot n > 0$. According to Ref.~\cite{Leibbrandt:1983pj}, the ML prescription is then given by, 
\begin{equation}
   \frac{1}{n \cdot q} = \, \lim _{\theta \rightarrow 0} \frac{n^*\! \cdot q}{\left(n^*\! \cdot q\right)(n \cdot q)+i \theta} \, .
\end{equation}
where $\theta >0$. In an integral the prescription leads to,
\begin{align}
  \int d^4 q & \, \frac{n^*\! \cdot q}{\left(n^*\! \cdot q\right)(n \cdot q)+i \theta} \nonumber \\
  =   \int d^4 q & \,  \frac{n^*\! \cdot q}{n_0^2\left(q_0^2-\frac{(\vec n \cdot \vec q \,)^2}{n_0^2}+\frac{i \theta}{n_0^2}\right)} \ , 
\end{align}
from which can be read that the poles are located in the second and fourth quadrants,
\begin{equation}
   q_0= \frac{1}{n_0} \left ( \pm  |\vec n \cdot \vec q\, |  \mp i \frac{ \theta}{2 |\vec n \cdot \vec q\, |} \right )  \, .
\end{equation}
This allows for a Wick rotation using a counter-clockwise path realized with the Euclidian ML prescription,
\begin{equation}
    \left ( \frac{1}{n \cdot q} \right )_{\! E} =  \lim_{\theta^2 \to 0} \  \frac{(n^* \! \cdot q)_E}{(n^*\! \cdot q )_E\,  (n \cdot q)_E+\theta^2} \ ,  
\end{equation}
where $n_0 = in_4= in_4^*$, $q^2 = -q_E^2$ and $n\cdot q = - (n\cdot q)_E$. 

When a diagram includes gluon and quark propagators, the corresponding integral contains  three types of poles, all of which are located in the second and fourth 
quadrants of the complex plane. Therefore, in calculating the self-energy correction of the quark propagator at one loop, a Wick rotation to Euclidean space is 
straightforward. 

At one-loop order, the trace of the quark self-energy $\Sigma(p,n,n^*)$ leads to a tensor structure whose most general decomposition is given by the 
sum~\cite{Leibbrandt:1983zd,Mirja2020},
\begin{align}
\hspace*{-2mm}
   \Sigma(p,n,n^*) & =  a\,  i\gamma\cdot p + b\,\mathbb{1} +  c\, i \gamma\cdot n \nonumber \\
     & -  d\, i (  \gamma\cdot n \, \gamma\cdot n^* \gamma\cdot p  + \gamma\cdot p\ \gamma\cdot n^* \gamma\cdot n) \, ,
\label{DSEdecompose1}     
\end{align}
and all vectors and Dirac matrices are now in Euclidean space. The $d$-term stems from the regularization procedure and can be further reduced to amplitudes 
proportional to $p$, $n$ and $n^*$ using the identity,
\begin{align}
\hspace*{-1mm}
   \gamma\cdot n\, \gamma  & \cdot n^*  \gamma\cdot p   +\gamma \cdot  p\,  \gamma\cdot n^*  \gamma \cdot n      \nonumber \\
   = & \ (2\, n^*\! \cdot p ) \gamma \cdot n -  (2\, n \cdot p)  \gamma\cdot n^*\! + (2\, n^*\!\! \cdot n )  \gamma\cdot p \, .
\end{align}   
This simplifies Eq.~\eqref{DSEdecompose1}  which becomes:
\begin{equation}
     \Sigma(p,n,n^*) =    A\,  i\gamma\cdot p +  B\,\mathbb{1} +  C\, i \gamma\cdot n  +  D\, i \gamma\cdot n^* \, , \\ 
  \label{selfenergypert}   
\end{equation}
where all four dressing functions are functions of $p$, $n$ and $n^*$. 
The structure of Eq.~\eqref{selfenergypert} can be understood within the framework of the Newman-Penrose tetrad scheme~\cite{Newman:1961qr}, in which any
four-dimensional vector can be represented as a superposition of four null-vectors, and is discussed in detail in Ref.~\cite{Leibbrandt:1984be}.

Appropriate projections allow to obtain integral equations for all four amplitudes, $A, B, C$ and $D$, and in a one-loop calculation they can be solved with the usual 
Feynman parametrizations, though the four equations remain uncoupled. On the other hand, in a nonperturbative functional approach solving the DSE of the quark,  
the amplitudes are described by a set of nonlinear coupled integral equations. We solve this system of  equations in light-cone gauge and discuss their numerical 
solutions in the  following sections.


\section{Gap equation in light-cone gauge}

For the phenomenological reasons expounded in Section~\ref{intro}, we are interested in nonperturbative quark propagators which are solutions of the quark 
gap equation. To this end, we solve the equation of motion of the quark, which in quantum field theory is known as DSE. For the quark of a given flavor the inverse
quark propagator is obtained from\footnote{ \protect We employ Euclidean metric, $\left\{ \gamma_{\mu},\gamma_{\nu} \right\} = 2 \delta_{\mu\nu}$, with hermitian 
Dirac matrices: $\gamma^{\dagger}_{\mu} =\gamma_{\mu}$. Furthermore, $\gamma_{5}= \gamma_5^\dagger = \gamma_{4} \gamma_{1} \gamma_{2} \gamma_{3}$,
 with  $\mathrm{Tr}\left[ \gamma_{5} \gamma_{\mu} \gamma_{\nu} \gamma_{\alpha} \gamma_{\beta} \right] = -4 \epsilon_{\mu \nu \alpha \beta}$,
$\sigma_{\mu \, \nu} = \frac{i}{2} \left[ \gamma_{\mu},\gamma_{\nu} \right ]$ and a space-like vector $p_\mu$ is characterized by $p^2 >0$.}, 
\begin{align}
  S^{-1} (p)  & = \, Z_2 \,  i\, \gamma\cdot p + Z_4 \, m(\mu)   \nonumber \\ 
            + \, Z_1 &  g^2 \int^\Lambda\!\!  \frac{d^4k}{(2\pi)^4}\  D^{ab}_{\mu\nu} (q)\, \gamma_\mu t^a\,  S (k) \,\Gamma_\nu^{b} (k,p) \, ,     
  \label{DSEquark}                             
\end{align}
where $q=k-p$ is the gluon momentum, $Z_1(\mu,\Lambda)$,  $Z_2(\mu,\Lambda)$ and $Z_4(\mu,\Lambda)$ are the vertex, wave function and mass renormalization 
constants, respectively,  $m(\mu)$ is the renormalized current-quark mass and  $t^a = \lambda^a/2$ are the SU(3) group generators. In the self-energy integral, 
$\Lambda$ is a Poincar\'e-invariant cut-off  while  $\mu$ is the renormalization point chosen such that $\Lambda \gg \mu$. For a review of phenomenological
applications of the DSE, we refer to Ref.~\cite{Bashir:2012fs}.

The quark propagator is a gauge-dependent Green function and if we specify the light-front gauge, the gluon propagator must be of the form in Eq.~\eqref{LCgluon}. 
Generalizing, the nonperturbative gluon propagator can be written as~\cite{Cornwall:1981zr},
\begin{equation}
   D_{\mu \nu}^{a b} (q)  =    \delta^{a b}\,  \Delta  (q^2)   \left(\delta_{\mu \nu}-\frac{n_\mu q_\nu+n_\nu q_\mu}{n \cdot q}\right)  ,
 \label{LCgluonnonpert}  
\end{equation}
where  $\Delta(q^2)$ is the gluon's dressing function which we introduce in Eq.~\eqref{alpheff}. Taking into account Eq.~\eqref{selfenergypert}, the solutions of the 
DSE~\eqref{DSEquark} are generally written as,
\begin{equation}
    S^{-1} (p,n,n^*)   =   A \,  i\gamma\cdot p +  B\,\mathbb{1}   +  C\, i \gamma\cdot n  +  D\, i \gamma\cdot n^* . 
 \label{LCquark}        
\end{equation}
The propagator differs from its form in covariant gauges not merely because of two additional scalar amplitudes, but also due the dependence of \emph{all} 
scalar functions on the relative orientation of the four-vectors $p$, $n$ and $n^*$. We will see that this is a signature of the light-cone gauge dependence.  

The  quark-gluon vertex, $\Gamma_\mu^{a} (k,p) \equiv \Gamma_\mu (k,p) t^a$, has been the object of much attention for  the past two decades and was shown to 
be crucial for the enhancement of the strong interaction in the infrared domain, and thereby for dynamical chiral symmetry breaking 
and the \emph{emergence of a constituent mass scale\/}~\cite{Curtis:1990zs,Fischer:2003rp,Alkofer:2008tt,Kizilersu:2009kg,Williams:2014iea,Bashir:2011vg,
Bashir:2011dp,Rojas:2013tza,Rojas:2014tya,El-Bennich:2016qmb,Binosi:2016wcx,Bermudez:2017bpx,Serna:2018dwk,Albino:2018ncl,Albino:2021rvj,Lessa:2022wqc,
El-Bennich:2022obe,Aguilar:2010cn,Aguilar:2014lha,Aguilar:2024ciu}. This vertex must satisfy gauge invariance and current conservation, which in Abelian theory is 
imposed by a Ward-Fradkin-Green-Takahashi identity (WFGTI)~\cite{Ward:1950xp,Fradkin:1955jr,Green:1953te,Takahashi:1957xn} and in Yang-Mills theories by 
Slavnov-Taylor identities~(STI)~\cite{Slavnov:1972fg,Taylor:1971ff}.  

On the other hand, since the ghosts decouple in light-cone gauge, the sum of the abelian and non-abelian contributions to the quark-gluon vertex satisfy a 
WFGTI~\cite{Leibbrandt:1983zd}  and can be decomposed as,
\begin{align}
  \Gamma_\mu (k, p,n,n^*)  & =   \lambda_1 \gamma_\mu + \lambda_2 \, i \gamma\cdot n^*  n_\mu \nonumber \\ 
                                            & +  \,  \lambda_3\, i \gamma\cdot n\,  n_\mu^*   +   \lambda_4 \,  i\gamma\cdot n\, q\cdot  n^* n_\mu \, ,
\label{perturbvertex}                                            
\end{align}
where the form factors $\lambda_i \equiv \lambda_i (k,p,n,n^*)$ are known to one-loop~\cite{Leibbrandt:1983zd}. However, since the form of Eq.~\eqref{perturbvertex} 
is derived in perturbation theory, the corresponding Ward identity is valid for propagators with constant quark mass, but not for the nonperturbative solutions of
Eq.~\eqref{LCquark}.

We therefore generalize the vertex in Eq.~\eqref{perturbvertex} and in analogy with the standard Ball-Chiu vertex we add the two components,  
$\gamma\cdot (k + p)  (k+p)$ and $(k+p)$. Thus, combining the two terms proportional to $n_\mu$ in Eq.~\eqref{perturbvertex}, our ansatz is given by the non-transverse 
vertex decomposition, 
\begin{align}
    \Gamma_\mu (k,p,n,n^*)  & =   \lambda_1 \gamma_\mu + \lambda_2\, \gamma\cdot \left (  k +  p \right )  (k+p)_\mu   \nonumber \\
                                              -  i\lambda_3 & (k+p)_\mu   +  \lambda_4 \gamma \cdot n^* n_\mu +  \lambda_5  \gamma \cdot n \, n_\mu^*  \, ,
 \label{WT-vertex-LC}                                             
\end{align}      
which we insert in the WFGTI of the quark-gluon vertex,
\begin{equation}
   i q\cdot \Gamma (k,p,n,n^*) = S^{-1}(k,n,n^*) - S^{-1}(p,n,n^*)   \, ,
\label{WGTI}   
\end{equation}
along with the inverse quark propagator in Eq.~\eqref{LCquark}. The left-hand side of Eq.~\eqref{WGTI} becomes, 
 \begin{align}
    i  ( & k-p)  \cdot \Gamma (k,p,n,n^*)  \,  =  \lambda_3 (k^2-p^2)  \mathbb{1}    \\
           + & \ i\big [ \lambda_1 +  \lambda_2 (k^2-p^2)  \big  ]  \gamma\cdot k  + i \left [ \lambda_2 (k^2-p^2) - \lambda_1 \right  ] \gamma \cdot p 
            \nonumber \\
           + & \  i \lambda_4 \left ( n\cdot k - n \cdot p \right )  \gamma \cdot n^*  +  i \lambda_5 \left ( n^*\!\cdot k - n^*\! \cdot p \right ) \gamma \cdot n \, , 
           \nonumber
\end{align}
while the right-hand side is,
\begin{align}
    S^{-1} & (k , n,n^*)   -  S^{-1} (p, n,n^*) =   \nonumber \\   & \ A(k)\,  i\gamma\cdot k +  B(k)\,\mathbb{1} +  C(k)\, i \gamma\cdot n  +  D(k)\, i \gamma\cdot n^*    \nonumber \\
                       -  & \, A(p)\,  i\gamma\cdot p -  B(p)\,\mathbb{1} - C(p)\, i \gamma\cdot n  -   D(p)\, i \gamma\cdot n^*  =   \nonumber \\
                         &  \, \left [ B(k) - B(p) \right ]  \mathbb{1}  + i \left [ A(k)\, \gamma\cdot k - A(p)\,  \gamma\cdot p \,\right ]   \nonumber \\
                       +  & \, i  \left [ C(k) - C(p) \right ]  \gamma\cdot n  + i  \left [ D(k) - D(p) \right ]  \gamma\cdot n^* \, . 
\end{align}
Equating same Dirac structures on both sides leads to the form factors:
\begin{align}
   \lambda_1  (k,p,n,n^*) & =  \frac{A(k)) + A(p,))}{2} \ ,  \\
   \lambda_2  (k,p,n,n^*) & =  \frac{1}{2} \frac{A(k)) - A(p)}{k^2-p^2} \ , \\
   \lambda_3  (k,p,n,n^*) & =  \frac{B(k) - B(p)}{k^2-p^2} \ , \\
   \lambda_4  (k,p,n,n^*) & =  \frac{D(k) - D(p)}{n\cdot (k-p)}    \ , \\
   \lambda_5  (k,p,n,n^*) & =  \frac{C(k) - C(p)}{n^*\!\cdot (k-p)} \, , 
\end{align}
where a dependence of the scalar amplitudes $A, B, C$ and $D$ on the light-like vector $n$ and $n^*$ is implicit. 

If the gluon propagator and quark-gluon vertex were bare, the arguments presented in Section~\ref{sec2} that justify a Wick rotation would be valid, provided a constant 
quark-mass function with poles located in the second and fourth quadrants of the complex plane. This, however, is not the case as the nonperturbative mass function, 
defined by $M(p^2) = B(p^2)/A(p^2)$, is characterized  by complex-conjugate poles or branch cuts~\cite{El-Bennich:2016qmb}. An interpolation formalism from instant-frame
to light-front dynamics in a QCD  model in $1+1$  dimensions and infinite number of colors was shown to allow for a matching of Minkowski and Euclidean space~\cite{Ma:2021yqx}, 
though the case of QCD in  $3+1$ dimensions beyond perturbative studies is an open question. A detailed study of the singularity structure of propagators in the Bethe-Salpeter 
equation demonstrated that the light-front wave function calculated with light-front coordinates \emph{defined in\/} Euclidean metric is identical to the one in Minkowski metric 
for the case of a monopole model of the Bethe-Salpeter amplitude~\cite{Eichmann:2021vnj}, while the DSE in Minkowski space has been treated, for instance, in 
Refs.~\cite{Sauli:2002tk,Duarte:2022yur}. Eschewing  these conceptual difficulties, we here make bold to work directly in Euclidean space.

In order to solve the DSE~\eqref{DSEquark} the scalar amplitudes must be projected out with the following traces over color and Dirac indices:
\begin{align}
    A(p,n,n^*)   &  = -\dfrac{i}{4} \Tr \left [ \frac{\gamma \cdot n }{ n \cdot p} \, S^{-1} (p,n,n^*) \right ]  ,  \label{Aeqint}  \\
    B(p,n,n^*)   &  = \phantom{-} \frac{1}{4}  \Tr  \left [ S^{-1} (p, n, n^*)  \right ]  ,  \\
    C(p,n,n^*)  &  =  -\frac{i}{4}  \Tr  \left [ \frac{\gamma \cdot p }{ n \cdot p} \,  S^{-1} (p,n,n^*) \right ]  ,  \\
    D(p,n,n^*)  &  =  -\frac{i}{4}  \Tr  \left [ \frac{\gamma \cdot p }{n^*\! \cdot p} \, S^{-1} (p,n,n^*) \right ]  .  \label{Deqint} 
\end{align}

\noindent
With these projections we arrive at a set of coupled integral equations which is given  in rainbow truncation  by, 
\begin{widetext}
\begin{align}
 A (p,n,n^*) & =\, Z_2 + \frac{2 D(p,n,n^*)}{n \cdot p} - \frac{i Z_{2} g^{2} C_{f}}{4\, n \cdot p}\! \int^\Lambda\! \frac{d^4k}{(2\pi)^4}\, D_{\mu \nu}(q) 
                          \operatorname{Tr}_D \left[\gamma \cdot n\,  \gamma_\mu S (k,n,n^*) \gamma_\nu \right ] ,
  \\[0.6em]
 B (p,n,n^*) & =\, Z_4 m (\mu) + \frac{Z_{2}  g^{2} C_f}{4}\! \int^\Lambda\! \frac{d^4k}{(2\pi)^{4}}\,  D_{\mu \nu}(q) \operatorname{Tr}_D
                           \left[\gamma_\mu S (k,n,n^*) \gamma_\nu \right ] , 
  \\[0.4em]
 C(p,n,n^*) & = \frac{p^{2}}{n \cdot p} [ Z_{2} - A(p,n,n^*) ] - D(p,n,n^*)\frac{n^*\!\cdot p }{n \cdot p} - \frac{i Z_{2} g^{2} C_{f}}{4\, n \cdot p}  
                         \! \int^\Lambda\!\!\! \frac{d^4k}{(2\pi)^4}\, D_{\mu \nu}(q)  \operatorname{Tr}_D   [ \gamma \cdot p \, \gamma_\mu S (k,n,n^*) \gamma_\nu  ]\, ,  \label{CTr}
   \\[0.6em]                   
 D(p,n,n^*) & = \frac{p^{2}}{n^*\! \cdot p} [Z_{2} - A(p,n,n^*) ] - C(p,n,n^*)\frac{n\cdot p }{n^*\! \cdot p} - \frac{i Z_{2}  g^{2} C_{f}}{4\, n^*\!\cdot p} 
                        \!\! \int^\Lambda\!\! \! \frac{d^4k}{(2\pi)^4}\, D_{\mu \nu}(q) \operatorname{Tr}_D  [ \gamma \cdot p \, \gamma_\mu S (k,n,n^*) \gamma_\nu  ] \, .   \label{DTr}                                         
\end{align}
\end{widetext}
where $C_F = 4/3$ stems from the color trace and we use $Z_1=Z_2$, since the quark-gluon vertex satisfies the Abelian WFGTI~\eqref{WGTI}.  

Closer inspection of Eqs.~\eqref{CTr} and \eqref{DTr}, multiplying them by $n\cdot p$ and $n^*\!\cdot p$ respectively, reveals that they are identical. 
Indeed, one of the functions, either $C (p,n,n^*)$ or $D (p,n,n^*)$, is superfluous and it turns out that $D(p,n,n^*) \equiv 0$. This is not surprising, as the introduction 
of $n^*$ is an artifact to tame divergences stemming from the $(n\cdot q)^{-1}$ factor in the gluon propagator~\eqref{LCgluonnonpert}.  As we will see, the poles 
occurring in the denominators of the DSE~\eqref{DSEquark} can be dealt with differently. 

For the nonperturbative gluon propagator, that incorporates the correct ultraviolet behavior and is renormalizable,  we employ the propagator derived by Cornwall
within the framework of pinch technique~\cite{Cornwall:1981zr}, 
\begin{equation}
  \Delta^{-1} (q^2 ) = \left[ q^2 + m_g^2 (q^2) \right ] \! bg^2 \ln \left[\frac{q^2+4 m_g^2 (q^2) }{\omega^2} \right ] ,
\label{alpheff}                 
\end{equation}
with the running dynamical gluon mass,

\begin{equation}
   m_g^2 (q^2 )=  m_g^2 \left( \frac{\ln \left [(q^2+4 m_g^2) / \omega^2 \right ]  } { \ln \left [ 4 m_g^2 / \omega^2 \right ] } \right )^{\!\!-\tfrac{12 }{ 11}} \!\! .
\end{equation}
The relationship between $q$, $\omega$ and $\mu$ is defined by  $\Delta^{-1} (q^2 = \mu^2) = 1/\mu^2$ aside of mass terms, and we remark that the product 
$g^2 \Delta (q^2 )$ is independent of the strong coupling. The values for $b$, $m_g$ and $\omega$ will be discussed in Section~\ref{gaugedependence}.  
We remind Cornwall's argument that the full, renormalized propagator should remain a function of Lorentz-invariant quantities only, and that in light-cone gauge 
a scalar Green function of one momentum cannot depend on $n_\mu$~\cite{Cornwall:1981zr,Cornwall:1974hz}; 
thus, the dressing function $\Delta (q^2 )$ only depends on $q^2$, whereas the light-like vector appears in the free propagator~\eqref{LCgluon}.  

We note that a contemporary understanding~\cite{Binosi:2014aea} of the pinch technique led to the development of a renormalization-group-invariant (RGI)
running  interaction, computed via a combination of DSE- and lattice-QCD results in Ref.~\cite{Aguilar:2009nf}. We employed both, the RGI interaction and
the Cornwall propagator in Eq.~\eqref{alpheff}, to solve the integral equations~\eqref{Abareeq}, \eqref{Bbareeq} and \eqref{Cbareeq}, and found minimal 
qualitative and quantitative differences for the scalar functions $A(p,n)$, $B(p,n )$ and $C(p,n)$. The numerical results presented in Sec.~\ref{gaugedependence}
were obtained with the pinch-technique gluon propagator of Eq.~\eqref{alpheff}. 

After taking the Dirac traces, we are left with three coupled integral equations, 
\begin{align}
  A(p,n )   = & \  Z_2   
                  + \frac{2 Z_2 C_f}{ n\cdot p} \! \int^\Lambda\!\! \frac{d^4k}{(2\pi)^4}  g^2 \Delta(q^2) \,  \sigma_{A}(k,n) \,  n\cdot k \, ,  
                     \label{Abareeq}    \\
  B(p,n )  = &\  Z_4 m (\mu) \nonumber \\ 
                     + & \ 2 Z_2 C_{f}  \int^\Lambda\! \frac{d^4k}{(2\pi)^{4}}  g^2 \Delta(q^2)\, \sigma_{B}(k,n)\, ,  \label{Bbareeq}    \\
  C(p,n)  = &  \ Z_c + \frac{p^{2}}{n \cdot p} \big  [ Z_2 - A(p,n) \big ] 
  \nonumber \\
    + & \ 2 Z_2  C_f  \int^\Lambda\! \frac{d^4k}{(2\pi)^4} g^2 \Delta(q^2)  \Big [  \sigma_C(k,n)   \nonumber \\
                   + &   \left.    \sigma_{A} (k,n )  \frac{ n\cdot p \;  k\cdot q+ n\cdot k \ p\cdot q }{n\cdot p \; n \cdot q}    \right ] , 
  \label{Cbareeq}
%
\end{align}
where we define the scalar dressing functions,
\begin{align}
\sigma_{A}(k,n) =  & \  \frac{A(k,n)}{ \mathcal{F}(k,n)  } \, ,  \\
\sigma_{B}(k,n) = & \ \frac{B(k,n )}{ \mathcal{F}(k,n) } \, ,   \\
\sigma_{C}(k,n) = & \ \frac{C(k,n)}{ \mathcal{F}(k,n) } \, , 
%
\end{align}
and the denominator is, 
\begin{align}
   \mathcal{F}(k,n)  = & \  k^2 A^2(k,n)  + B^2(k,n)  \nonumber \\
    & \ + 2A(k,n )\, C(k,n)\, n\cdot k  \, .
\end{align}
These coupled integral equation are solved iteratively, imposing the renormalization conditions,
%
\begin{figure}[t!] 
\centering
  \includegraphics[scale=0.74,angle=0]{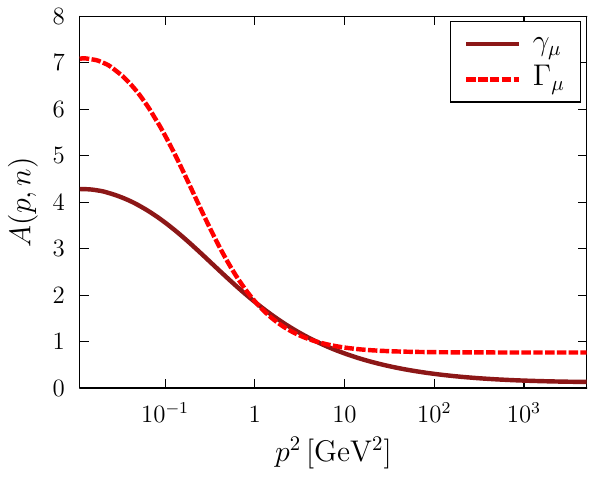}  \\
  \includegraphics[scale=0.74,angle=0]{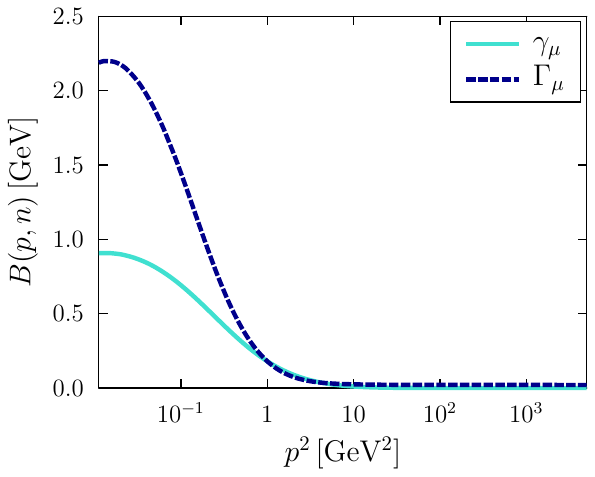} 
\caption{The solution $A(p,n)$ and $B(p,n)$ of the quark DSE obtained with  a bare vertex and with the WFGTI ansatz of Eq.~\eqref{WT-vertex-LC}.}
\label{figA-B} 
\end{figure}
%
\begin{figure}[t!] 
\centering
  \includegraphics[scale=0.74,angle=0]{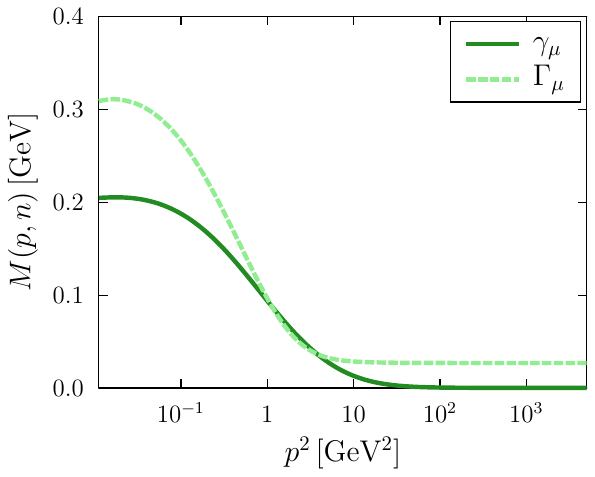} 
\vspace{-1mm}
\caption{The quark-mass function $M(p,n) = B(p,n)/A(p,n)$ obtained with the bare and WFGTI vertices. }
\label{figM} 
\end{figure}
%
\begin{equation}
    B (\mu,n) =  m(\mu) = 25\; \mathrm{MeV} \quad \text{and} \quad A (\mu,n) =1 \ ,
\end{equation}
at the renormalization scale $\mu = 2$\;GeV, and we introduce the renormalization constant $Z_c$ in Eq.~\eqref{Cbareeq}  to the effect that $C(\mu,n) = 0$ at small momenta, 
as suggested by a one-loop calculation  in Ref.~\cite{Mirja2020}. The cut-off can be formally taken to infinity, though numerically we employ $\Lambda =100$\;GeV and our
DSE solutions are stable for larger values of $\Lambda$. Likewise, we derive the integral equations~\eqref{Aeqint} to \eqref{Deqint} for the dressed vertex ansatz in 
Eq.~\eqref{WT-vertex-LC}, the expressions of which we here omit  due to their length.  

\begin{figure}[t!] 
\centering
  \includegraphics[scale=0.8,angle=0]{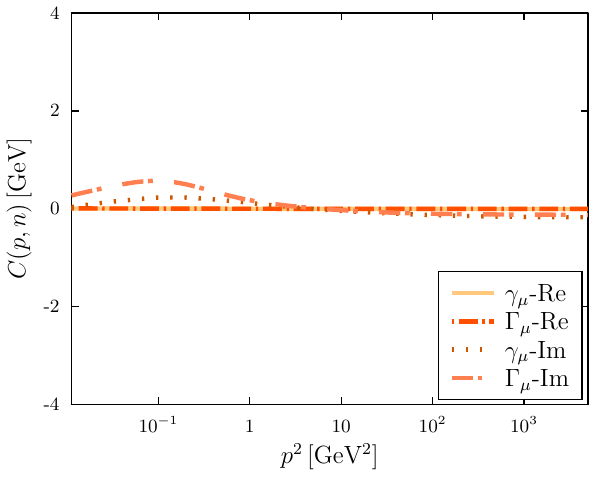}
\caption{Real and imaginary parts of the dressing function $C(p,n)$ obtained with the bare and WFGTI vertices.}
\label{figC} 
\end{figure}


\section{Gauge dependence of the quark propagator}
\label{gaugedependence}

We compare numerical solutions of the quark's dressing functions for both, the bare vertex and the vertex in Eq.~\eqref{WT-vertex-LC}, that is the solutions of 
Eqs.~\eqref{Abareeq}, \eqref{Bbareeq} and \eqref{Cbareeq} in the \emph{rainbow} approximation and their counterpart using the WFGTI-vertex. 
In both cases we use the values,  $m_g =0.3$~GeV and $ \omega = 0.15$~GeV, in Eq.~\eqref{alpheff}, while $b = 2.0 \times 10^{-4}$ for the bare vertex and
$b = 9.0 \times 10^{-3}$ for the WFGTI-vertex. The dressing functions $A(p,n)$ and  $B(p,n)$ are presented in Fig.~\ref{figA-B} in the quark's rest frame,  
$p = (\mathbf{0}, |p| )$, from which we infer that their functional behavior is  reminiscent of that found in covariant gauges~\cite{Lessa:2022wqc}.  In Fig.~\ref{figM} one 
observes the typical rapid increase of the mass function at a hadronic scale of about 1\,GeV and  $M(0) \approx 210$\;MeV when the bare vertex is employed. 
Decreasing the value of $b$ increases the mass function, though the solutions are unstable. We choose the parameters, $m_g$ and $\omega$, to be the same
in both cases, however, slight modifications of them results in a constituent mass above 300\;MeV even when solving the DSE with the bare vertex. Unlike the two 
other dressing functions $C(p,n)$ is complex valued, as illustrated in Fig.~\ref{figC}. Both the real and imaginary parts of  $C(p=\mu, n)$ are zero due to our 
renormalization prescription, and one merely observes insignificant oscillations of these function consistent with  $C(p,n) \approx 0$ in the entire range of $p^2$ considered.

\begin{figure}[t!] 
\centering
  \includegraphics[scale=0.79,angle=0]{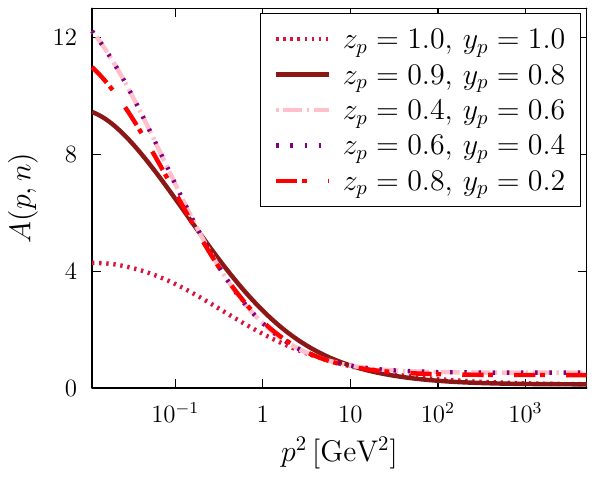} \\  \vspace*{1mm}
  \includegraphics[scale=0.79,angle=0]{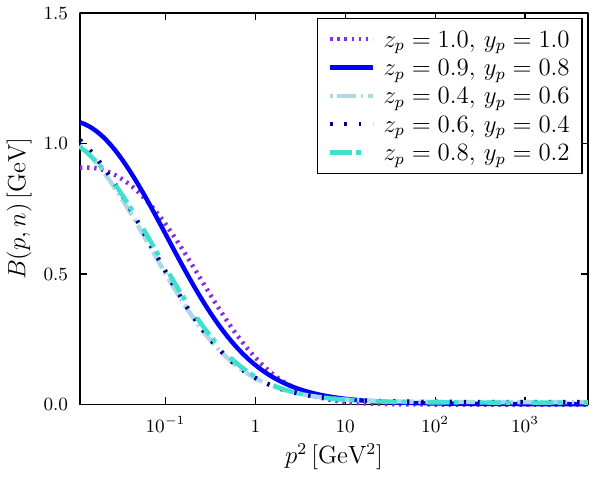} \\
\caption{The gauge dependence of the scalar functions, $A(p,n)$ and $B(p,n)$, obtained with a bare quark-gluon vertex is due to the relative orientation of the 
vectors $n$ and $p$ and expressed by the angles $y_p$ and $z_p$ in Eq.~\eqref{anglep}.}
\label{figA-B-angledep} 
\end{figure}
%
\begin{figure}[h!] 
\centering
  \includegraphics[scale=0.79,angle=0]{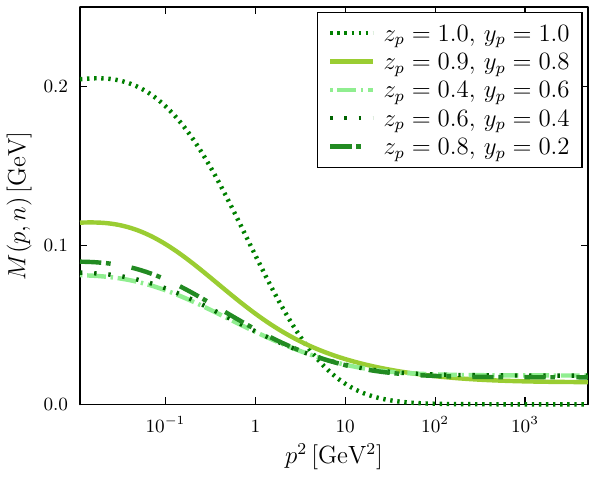} 
\caption{The dependence of the quark-mass function $M(p,n) = B(p,n)/A(p,n)$ on the quark momentum $p$ in light-cone direction, see Eq.~\eqref{anglep}. }
\label{figM-angledep} 
\end{figure}

Turning to the case of the dressed vertex that satisfies the WFGTI~\eqref{WGTI}, the integral equations for $A(p,n)$, $B(p,n)$ and $C(p,n)$ are more lengthy and the 
convergence of the numerical iterative procedure is considerably slower. So as to produce a constituent mass that is comparable with that of  the bare vertex, 
we readjust the gluon dressing with $b = 9.0 \times 10^{-3}$, as already mentioned. The $A(p,n)$ and $B(p,n)$ functions are enhanced compared with those of the 
bare vertex in Fig.~\ref{figA-B}. While the mass functions are qualitatively alike below the renormalization point, $\mu =2$\;GeV, with a steeper rise in case of 
the WFGTI vertex, the behavior is different at larger momenta:  the mass function continuously decreases when a  bare vertex is employed, whereas it saturates at 
about 20\;MeV with the WFGTI vertex.  Moreover,  in Fig.~\ref{figC} we observe again that $C(p,n) \approx 0$.

As mentioned after Eqs.~\eqref{CTr} and ~\eqref{DTr}, for either vertex we find the numerical solution $D(p,n,n^*)= 0$. The inverse of the quark propagator can
therefore simply be written as:
\begin{equation}
     S^{-1} (p,n)   =   A \,  i\gamma\cdot p +  B\,\mathbb{1}   +  C\, i \gamma\cdot n\,  .
\end{equation}
For convenience, the solutions presented in Figs.~\ref{figA-B}, \ref{figM} and \ref{figC} were obtained in the rest frame of the quark.
In order to verify the dependence on the orientation of the quark momentum with respect to the light-like vector $n= (0,0,1,i)$, we solve the 
DSE  for an arbitrary momentum $p$. The scalar product of both vectors is then defined as, 
\begin{equation}
   n \cdot p = |p| \left (\sqrt{1-z_p^2 }\, y_p + i z_p \right )  ,
 \label{anglep}  
\end{equation}
with angles $y_p =\cos\theta_p$  and $z_p = \cos\psi_p$. This allows us to investigate how the three dressing functions of the quark propagator 
explicitly depend on the relative orientation of the four-vectors $n$ and $p$. A rigorous calculation requires that $A(k,n)$, $B(k,n)$ and $C(k,n)$
in the kernels of Eqs.~\eqref{Abareeq}, \eqref{Bbareeq}  and \eqref{Cbareeq} be functions of the three variables $p$, $y_p$ and $z_p$. Therefore, 
in solving iteratively this system of coupled integral equations, the values of $A(p,n)$, $B(p,n)$ and $C(p,n)$ must be obtained on a three-dimensional 
mesh for discrete values $p^i$, $y_p^k$ and $z_p^l$ in a first iteration and then fed back into the integral equations. As this is an exploratory study, 
we limit ourselves to the angular dependences of the explicit $n\cdot p$ terms in Eqs.~\eqref{Abareeq}, \eqref{Bbareeq} and \eqref{Cbareeq}.
For our purposes, this suffices to demonstrate the dependence on the orientation of the quark momentum relative to the light-like vector $n$. 
We note that the denominator of the third term in Eq.~\eqref{Cbareeq} possesses a characteristic $(n\cdot q)^{-1}$ pole when $z_k = +1$ and
$y_k = 0$, where the angles are between the vectors $n$ and $k$.  In treating the integral numerically, we limit ourselves to $-1 \leq z_k \leq +1 -\varepsilon$ 
with $\varepsilon \approx 10^{-3}$ and use the Cauchy principal value about the value $y_k=0$.

The behavior of the wave-renormalization and mass functions as functions of  $y_p$  and $z_p$ in the  bare-vertex truncation is illustrated in Figs.~\ref{figA-B-angledep} 
and \ref{figM-angledep} for a sample of representative angles. We observe an increase of $A(p,k)$ in the infrared domain $p^2\lesssim 1\;\text{GeV}^2$
while the variation in $B(p,k)$ is less marked when the angles depart from their initial values $y_p = z_p =1\, \Rightarrow\, n \cdot p =   i |p|$. Consequently, $M(p,k)$ is 
also suppressed for decreasing $z_p$. Note that $A(p,n)$ and $B(p,n)$ are identical for pairs of angles, $y_p, z_p$ and $y_p' = z_p, z_p' = y_p$, considering only
the explicit $n\cdot p$ terms. 

\begin{figure}[t!] 
\vspace*{-5mm}
\centering
 \includegraphics[scale=0.62,angle=0]{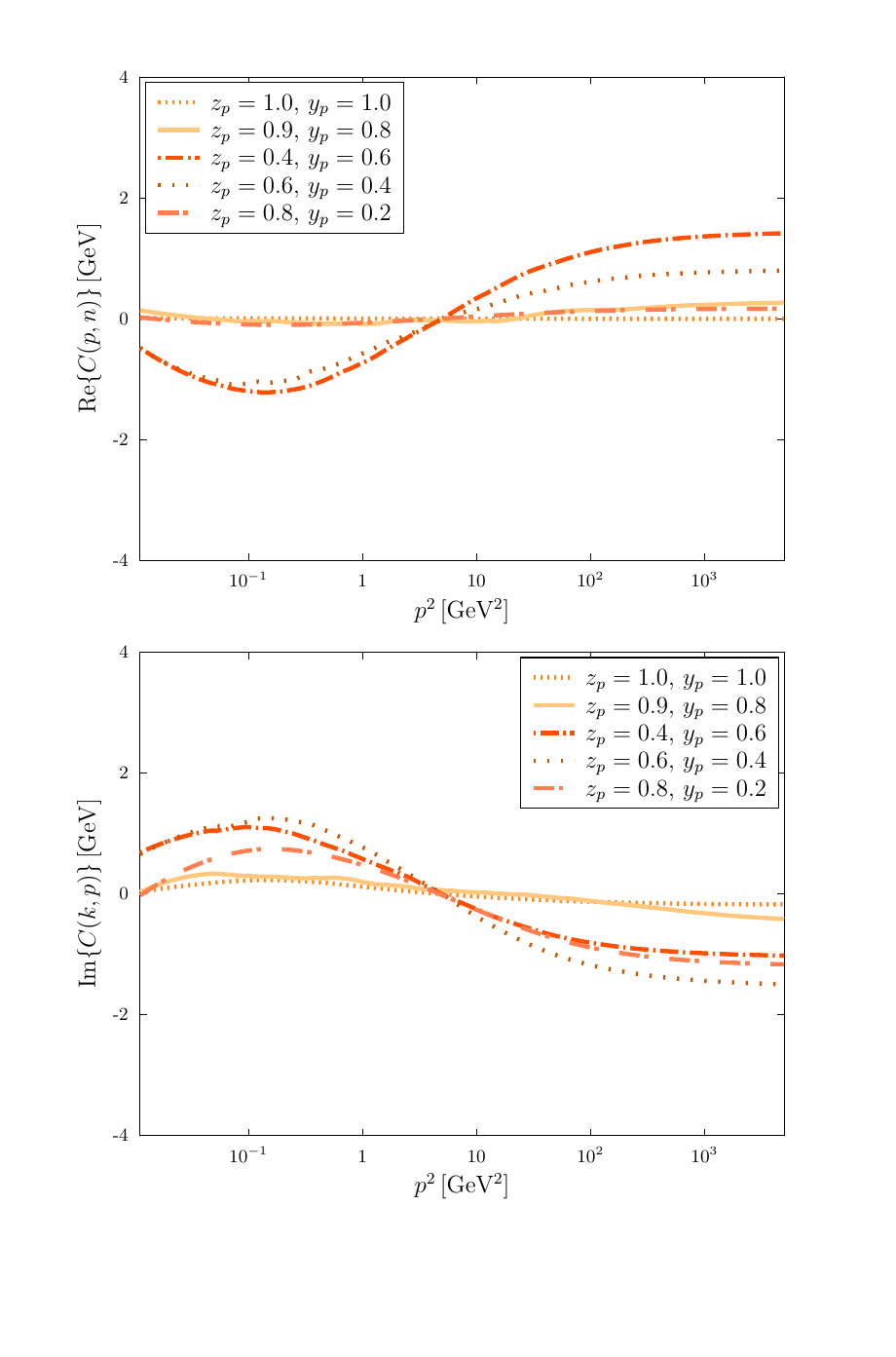} 
\vspace{-1.8cm}
\caption{Light-cone gauge dependence of the real and imaginary parts of the dressing function $C(p,n)$ calculated with a bare vertex.}
\label{figC-angledep} 
\end{figure}

The $n$-dependent variation of the complex dressing function $C(p,n)$ for the same set of angles is depicted in Fig.~\ref{figC-angledep}. It appears that when 
$y_p \neq 1$  and $z_p \neq 1$ the real and imaginary parts of $C(p,n)$ slightly oscillate about the renormalization point $\mu =2$\;GeV. We verified that these 
are not simply numerical fluctuations in the integration using both a deterministic cubature rule and a Monte Carlo method, namely the  {\tt CUHRE} and  {\tt VEGAS} 
algorithms of the {\tt CUBA} library~\cite{Hahn:2004fe}. At this point, not including the full $n$-dependence in the integral~\eqref{Cbareeq}, it is difficult to decide whether
$C(p,n) = 0$ for all momenta and possible angles. 

The convergence of the quark DSE with the vertex ansatz in Eq.~\eqref{WT-vertex-LC} is considerably slower for the values $z_p < 1$ and $y_p <1$ and solutions 
become unstable. We therefore abstain from presenting variations of $y_p$  and $z_p$ for this case.


\section{Concluding remarks}

Motivated by phenomenological considerations, namely the intricacy of treating a Wilson line between two quark fields in nonperturbative functional approaches without 
resorting to expansions or simply neglecting the operator in the definition of distribution functions, we made a first step towards a meson's wave function in light-cone gauge. 
To that end, we solved for the first time the DSE of a quark  in light-cone gauge employing Cornwall's gluon propagator~\cite{Cornwall:1981zr}. We showed that the ML 
prescription is moot in solving the relevant DSE and that the dual vector $n^*$ is not needed. 

As a consequence, the numerical solutions of the nonperturbative quark propagator contain three scalar functions, two of them playing the usual role of the mass 
and wave-renormalization functions, while a third dressing function is complex. The dependence of $A(p,n)$ and $B(p,n)$ on the relative orientation of the quark 
momentum $p$ in light-cone direction is found to be only significant below 1\;GeV, and the mass function is suppressed when moving away from the 
quark's rest frame. On the other hand, the real and imaginary parts of $C(p,n)$ are consistent with zero in the quark's rest frame, whereas in a moving frame
 $C(p,n)$ oscillates about the renormalization point. In other words, the breaking of covariance in light-cone gauge is manifest in the scalar functions.   
 
Naturally, a rigorous discussion of the quark gap equation in light-cone gauge requires the self-consistent solutions of the quark and the gluon in the same 
DSE framework and with their full dependence on the vector $n$. This may or may not compensate the suppression of the mass function. According to Cornwall, 
the dressing function of the gluon propagator is a covariant scalar function that only depends on its momentum~\cite{Cornwall:1981zr}. One may extend this
reasoning to the quark propagator and insist  that fully and correctly renormalized scalar Green functions of one momentum should be covariant. Leibbrandt's 
regularization  prescription to handle $1/n\cdot q$ singularities in light-cone gauge, on the other hand, leads to expressions for $A(p,n)$, $B(p,n)$ and $C(p,n)$ that 
do depend on the light-like vector, $n$, and in Ref.~\cite{Mirja2020} his calculation was  extended to include divergent terms renormalized in the $\overline{\text{MS}}$ 
scheme. Nevertheless, at the level of individual Feynman diagrams the singularities can lead to gauge-dependent terms proportional to the four-vector $n$ in self-energy 
corrections. Ultimately, these dependences might cancel.  Whichever it is, the quark propagator is not an observable and gauge dependent. Therefore, even 
if $C(p,n) \equiv 0$ was a true statement, the gauge dependence  should manifest itself in the wave-function renormalization and mass function. At this point, 
we cannot answer the question of how the quark propagator depends on the light-like vector.

Since in light-cone gauge the gluon decouples from the ghost, we also derived a dressed quark-gluon vertex that satisfies its WFGTI. With a readjustment 
of the gluon-dressing function we find that the dressing functions  $A(p,n)$,  $B(p,n)$ are suppressed in comparison with the same functions in the bare-vertex
truncation of the DSE, nonetheless the resulting mass function is qualitatively and quantitatively equivalent. Since the convergence of the iterative procedure
using Newton's method is very slow in case of the WFGTI vertex for arbitrary values of $n\cdot p$, we refrained from studying the mass function's dependence 
on it in this case. We remind that future improvements should implement the full angular dependence in the integral equations for $A(p,n)$, $B(p,n)$ and  $C(p,n)$.

With regard to the gauge independence of physical observables, only a symmetry preserving Bethe-Salpeter kernel for quark-antiquark states in the
same light-cone gauge framework can offer a sensible path. This difficult task has been pursued in Landau gauge and ought to be investigated in 
light-cone gauge in order to be able to calculate mesonic distribution amplitudes and parton distribution functions.


\acknowledgments

We would like to express our gratitude to Peter Tandy who suggested the study of the nonperturbative quark propagator in light-cone gauge and who has been of great help 
to interprete our findings.  We also appreciated stimulating discussions with Arlene Cristina Aguilar and Gast\~ao Krein. 
B.\,E. and R.\,C.\,S participate in the Brazilian network project \emph{INCT-F\'isica Nuclear e Aplica\c{c}\~oes\/}. This work was supported by the S\~ao Paulo Research 
Foundation (FAPESP), grant no.~2023/00195-8 and by the National Council for Scientific and Technological Development (CNPq), grant no.~409032/2023-9. R.\,C.\,S  
is supported by CAPES fellowship, grant no.~88887.086709/2024-00.


\end{document}